# Evaluating North American Electric Grid Reliability Using the Barabási-Albert Network Model

David P. Chassin[*], Christian Posse

*Pacific Northwest National Laboratory, Richland, WA 99352, USA*

**Abstract**

The reliability of electric transmission systems is examined using a scale-free model of network topology and failure propagation. The topologies of the North American eastern and western electric grids are analyzed to estimate their reliability based on the Barabási-Albert network model. A commonly used power system reliability index is computed using a simple failure propagation model. The results are compared to the values of power system reliability indices previously obtained using standard power engineering methods, and they suggest that scale-free network models are usable to estimate aggregate electric grid reliability.
© 2005 Elsevier Science B.V. All rights reserved.

**1. Introduction**

Extensive topological analysis of a wide variety of networks has yielded profound theoretical insights into the general characteristics of the systems they support and how they evolve. Generally, two classes of networks are recognized. Among the earliest identified were exponential networks, such as the random graphs described by Erdös and Rényi [1], which evolve to have a connectivity probability $P(k)$ that peaks at an average scale $\langle k \rangle$ and decays exponentially for large $k$. Watts and Strogatz [2] described small-world networks, which also lead to relatively homogenous topology. However, Barabási and Albert [3] discovered a class of inhomogeneous networks, called scale-free networks, characterized by a power-law connectivity probability $P(k) \propto k^{-\gamma}$. While highly connected nodes are improbable in exponential networks, they do occur with statistically significant probability in scale-free networks. Furthermore, the work of Albert *et al*. [4] suggests that these highly interconnected nodes appear to play an important role in the behavior of scale-free systems, particularly with respect to their resilience.

---

[*] *Correspondence address*: PO Box 999, MS K5-16, Richland, WA 99352. *E-mail address*: david.chassin@pnl.gov

In spite of publicity surrounding recent cascading power outages, the two major North American electric power grids are like other evolved complex systems described by Albert *et al*. in [4], i.e., they show a great degree of resilience to naturally occurring failures. In the case of electric transmission networks, the resilience to unusually high local failure rates is highly desirable. However, any prediction regarding the reliability of electric systems based on the Barabási-Albert model requires confirmation based on direct evidence that the systems and models used by electric system planners are indeed scale-free and that the reliability measurements observed by electric system operators are similar to those predicted using common failure propagation models.

Analyses of electric grid reliability using scale-free network models were initially reported by Watts and Strogatz in 1998 [2]. Based on a model of the western U.S. grid containing 4,941 nodes, they estimated $\gamma = 2.67$. More recently Albert, *et al*. [5] used the Platts database, which contained a total of 14,099 nodes (1,633 of which are power plants, 2,179 of which are distribution substations) and treated the North American power grid as a single system. They found that the cumulative degree distribution was proportional to an exponential, expressing a degree cut-off for nodes with large numbers of incident edges. This exponential cut-off is consistent with the impact of constraints described by Amaral *et al*. [11] and formulated by Newman *et al*. [12] as an additional exponential term such that $P(k) \propto k^{-\gamma} e^{-k/\kappa}$. The coefficient reported Albert *et al.* [5] in was $\kappa = 2$.

In this analysis we study the North American electric grid using two reliability study models for the western system and eastern interconnect, obtained from the Western Electricity Coordinating Council (WECC) and North American Electric Reliability Council (NERC), respectively. The advantages of using these models are significant if our purpose is to use the scale-free network models to better quantify grid resilience. First, these models represent the systems in question very accurately, operating in a "live" configuration, and under well-known outage study conditions. Second, these models contain all the assets involved in the reliability study. The Platts database is assembled from various reports, such as the so-called "Form 1" filed by public utilities with the Federal Energy Regulatory Commissions (FERC), and other non-reliability-related sources. It is important to note that not all utilities are required to report all their assets, so the Platts database is not used by the power industry as the basis for reliability studies. Third, we study the two networks separately (we excluded Texas, which is topological relatively isolated from the rest of North American). Considering the two models jointly as in the Platts database, does not necessarily reflect the nature of failure propagation: the interchange between systems is by design extremely small and via direct-current links only, so cascading outage propagation across the "seam" is considered highly unlikely, if not impossible.

**2. Electric Network Reliability Metrics**

Electric transmission network reliability indices are used to characterize and assess the robustness of bulk power systems and the reliability of power delivery at the customer supply-point. Bulk transmission systems exhibit a distinguishing characteristic that severe disturbances can have widespread impacts when they arise. Reppen and Feltes [6] identify three important characteristics of reliable bulk power systems: 1) low risk of widespread failure, 2) containment of failure when it does occur, and 3) rapid restoration of service. Following the 1965 Northeast blackout and the establishment of NERC, design criteria were developed for U.S. electric transmission systems to reduce the risk of such a cascading system failure. One result of this was the introduction of deterministic test criteria that include a variety of requirements evaluated on models of bulk transmission systems. However, a consequence of the computational complexity of bulk power systems simulation is that only a small subset of all possible cases are examined. Therefore, this subset is carefully chosen for the degree to which cases can lead to critical problems in the system. To further enhance the speed of these simulations, modelers of the transmission networks simplify many aspects of the topology that are invariant across the cases chosen.

Detailed system reliability simulations consider both system adequacy (defined as the ability of the system to supply demanded power) and security (defined as the ability of the system to operate normally in the event of disturbances). Models such as those used by Billington and Khan [7] address the degree to which both adequacy and security are satisfied by studying the large number of NERC-mandated failure cases. In general, simulations based on these types of models curtail energy consumption in response to increasing network constraints. At the end of the simulations for all cases, a number of system reliability indices are computed to determine the degree to which electric demand is met.

Supply point reliability indices address the local impact of bulk transmission outages by quantifying the frequency, duration, and severity of interruptions at the point of delivery. Generally, both deterministic and probabilistic assessments are performed and measures of reliability are estimated. According to Melo *et al.* [8], specific indices used to measure system reliability are the loss of load expectation (LOLE), loss of load probability (LOLP), expected power not supplied (EPNS), and expected energy not supplied (EENS). The LOLP corresponds to the expected value of a load state function $P(x)$, where $P(x) = 1$ when the $x$ is a failure state and $P(x) = 0$ when $x$ is operational state. The EPNS corresponds to the function $E(x)$, the amount of power that is desired but not delivered. LOLE and EENS can be obtained by multiplying the LOLP and EPNS by the study duration respectively. The frequency of failure is denoted LOLF, and corresponds to the function $F(x)$ such that $F(x) = 0$ when in the operational state, and $F(x)$ is the sum of the transition rates between $x$ and all operational states that can be reached from $x$ in one state transition. Finally, loss of load duration (LOLD) is obtained by computing the ratio of LOLP/LOLF.

The proposed method of estimating system LOLP is based only on the scale-free topological characteristics of a transmission network. To perform this analysis we recognize that the models for simulations of the power grid represent mainly the high-voltage network topology and the boundary conditions for loads include a substantial radial infrastructure related to sub-transmission and distribution grids, as shown in Fig. 1. Transmission networks involve power lines operating generally above 200 kilovolts (kV), sub-transmission involves lines generally between about 50 and 200 kV, and distribution systems operate usually below 50 kV.

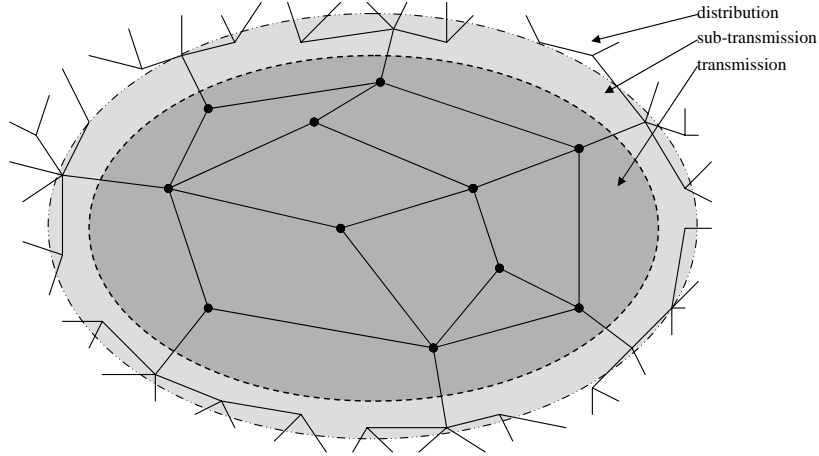

Fig. 1. The transmission network is a mesh system (dark gray), but sub-transmission (light gray) and distribution (white area) portions are mainly radial structures. Transmission system modeling places boundary nodes in the sub-transmission and distribution topology.

**3. Scale-Free Electric Networks**

According to the definition of a scale-free network, the number of $n(x)$ of nodes (referred to as buses by power systems engineers) having $x$ edges (referred to as branches) is given by

$$n(x) = n_1 x^{-\gamma} \qquad (1)$$

where $n_1$ is the number of nodes with only one edge (e.g., generators, nodes leading the radial load portions of the grid) and $\gamma$ is the power-law scaling parameter. We normalize this relation by dividing $n(x)$ by the total number of nodes $N$ in the system

$$\rho(x) = \frac{n(x)}{N} = \frac{n_1}{N} x^{-\gamma} = \rho_1 x^{-\gamma} \qquad (2)$$

where $\rho_1$ is the fraction of boundary nodes.

Topological analysis of the Eastern Interconnect [9] and Western System [10] electric transmission networks, shown in Fig. 2, confirms the scale-free nature of these transmission networks. However, we recognize that all single-edged nodes in the models in fact have two edges, the second of which is a connection either to a radial sub-transmission/distribution system or to a generator. As a result, the true value of $n_1$ is not directly observable from the available transmission system models. Fundamentally all remaining sub-transmission and distribution nodes must be connected to boundary nodes through the radial portion of the infrastructure.

We should also keep in mind that such topological studies do not consider that outage studies are performed on directed graphs representing "live" systems as opposed to the undirected graphs of "designed" systems. Simple connectivity statistics do not generally consider this distinction, in spite of the availability of the initial flow configuration in outage study models. The difficulty we confront in analyzing outage events is that flow reversals are not an infrequent occurrence. We will discuss below how aggregate models of failure propagation can overcome this problem without explicitly recomputing the power flow configuration of the directed graph.

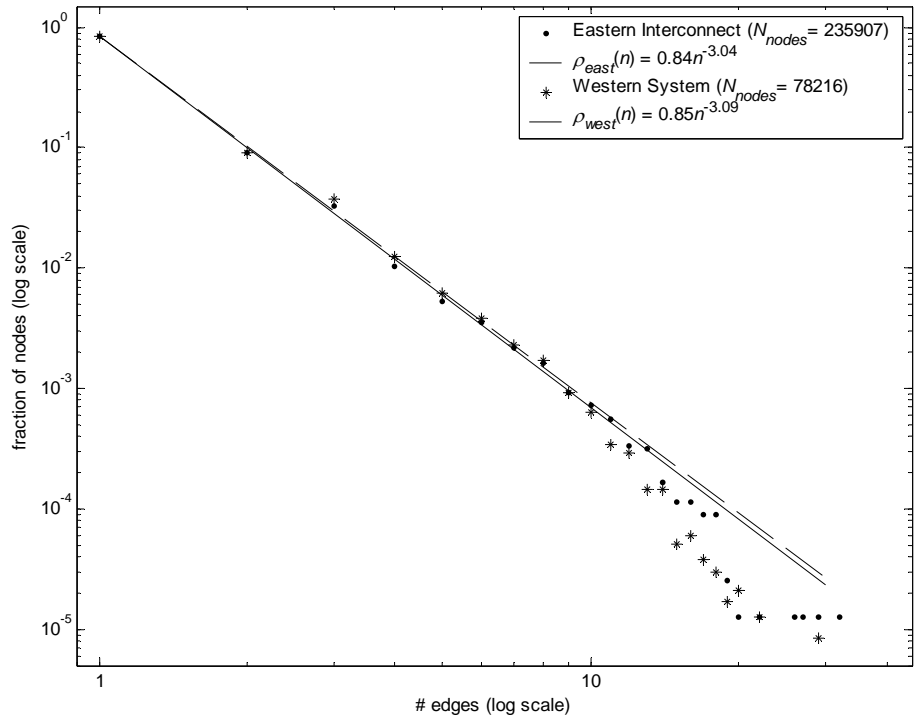

Fig. 2. Node (bus) connectivity statistics for the major North American electric grids.

Table 1. Estimates of effective number of nodes, loads, and generators on eastern and western systems

| System | Capacity | Feeder Phases (a) | Generators (b) | Network Nodes (c) | HV Grid (d) | $N=a+b+c$ | $N_1=N-d$ |
|---|---|---|---|---|---|---|---|
| Eastern Interconnect | 450 GW | 192,857 | 5,791 | 37,259 | 37,343 | 235,907 | 198,564 |
| Western System | 150 GW | 64,285 | 2,264 | 11,667 | 11,764 | 78,216 | 66,452 |

The problem of determining $\rho_1$ is addressed by using the estimates shown in Table 1 for the two network models considered. Boundary nodes are connected to either feeders or generators. Feeders are used to deliver three-phase electricity to customers on distribution systems. The number (a) of feeders is based on the assumption that the typical distribution feeder serves about 7 MW[a] of load distributed across all three phases[b]. The number (b) of generators is observed directly in the models. The total number of nodes $N$ is the sum of the number of feeder nodes, generator nodes, and network nodes (c) observed in the transmission system models obtained. The number of boundary nodes $N_1$ is the total number of nodes less the number of high-voltage nodes (d).

The parameter $\gamma$ was obtained by taking the logarithm of Eq. (2) and performing a least squares regression such that

$$\log \overline{\rho(x)} = -\gamma \log \bar{x} + \log \rho_1 \quad . \tag{3}$$

---

[a] The 7 MW feeder load estimate is based on the typical feeder load factor of 35% and maximum feeder capacity of 20 MW. [Kannberg *et al.*, "GridWise™: The Benefits of a Transformed Energy Systems", *PNNL Report No. 14396* (2003).]
[b] Switchgear control laws require that any failure on one phase or between two of the phases results in removal of all three phases. Therefore, they are treated as three separate points of origin but a single point of propagation.

Table 2: Topological properties of North American electric systems.

| Electric Grid | Total System Nodes | $\rho_1$ | $\sigma_\rho$ | $\gamma$ | $\sigma_\gamma$ |
|---|---|---|---|---|---|
| Eastern Interconnect | 235,907 | 0.84 | +0.092 / −0.083 | 3.04 | 0.059 |
| Western System | 78,216 | 0.85 | +0.067 / −0.062 | 3.09 | 0.047 |

According to Amaral *et al.* [11] deviations in power-law scaling are explained by cost and capacity constraints. Such a cut-off was formalized by Newman *et al.* [12] by adding the exponential $\kappa$ term to Eq. 2. To avoid the effect of this exponential cut-off, the averages were gathered only for those nodes that had 10 or fewer edges, ignoring those very few nodes that can approach cost or capacity constraints. The values obtained from this analysis are shown in Table 2.

Analysis of the available data confirms that both major North American electric transmission networks are Barabási-Albert scale-free networks, with the characteristic cut-off described by Amaral *et al*. System expansion costs tend to favor the use of existing facilities up to the point that their capacities are reached. The larger deviation of the Eastern Interconnect suggests that it is somewhat more constrained and expansion is more costly than in the Western System, perhaps simply attributable to the greater density of the eastern network. This in turn would lead one to expect higher asset utilization on the Eastern Interconnect than in the Western System.

## 4. Failure Propagation Model

Given an arbitrary component failure, resulting in either the removal of a node or edge, we can determine the LOLP by integrating the product of the connectivity probability and the propagation probability over all propagation path lengths for all *N* nodes. Therefore, to confirm the reliability predictions based on a scale-free network model, we propose a failure propagation model that is consistent with theory of operation of the transmission network.

Fundamentally there are two component-type failures that contribute to cascading failures in electric networks: edge removal (e.g., a line fails to maintain the fully energized condition of the bus[a] it is supporting) and node removal (e.g., a bus is itself under or de-energized). In the event of an edge removal, we reason that the probability of a node also being removed is primarily dependent on the number of edges connected to it. If a node has only one edge, then when that edge is removed, the probability of that node also being removed is 1.0, so the system aggregate probability of a single-edge node failing is 1.0 x $\rho_1$. However, as the number on a node of edges increases, the propagation probability drops significantly. For nodes with two edges, the system aggregate probability of node removal is roughly 1/2 x $\rho(2)$, and with three edges it is 1/3 x $\rho(3)$, etc. Thus, the discrete form for the system aggregate probability of an edge removal propagating to and causing node removal is

$$P\{\to node\} = \sum_{x=1}^{\infty} \frac{1}{x} \rho(x) = \sum_{x=1}^{\infty} \frac{\rho_1 x^{-\gamma}}{x} = \sum_{x=1}^{\infty} \rho_1 x^{-\gamma-1} \qquad (4)$$

---

[a]  *Bus* is the power engineering term for a node at which various kinds of connecting equipment arrive. Loads (located at feeders) and generators are always connected to a bus. An edge is referred to as a *branch*, and can include lines (low or high voltage), transformers (which convert to and from low and high voltages), and other equipment that interconnect buses.

The continuous form of (4) gives us the limit for a system of infinite[a] size

$$P\{\to node\} = \int_1^{\infty} \rho_1 x^{-\gamma-1} dx = \frac{1}{\gamma} \rho_1. \qquad (5)$$

The probability $P\{\to edge\}$ that a node removal propagates to an edge is simpler to determine because of the directed nature of power flow in the network. There are two possible relationships between a node and an incident edge, only one of which permits propagation from node to edge. Either power flows to the node, and the edge supports the node (i.e., propagation unlikely), or power flows from the node and the node supports the edge (i.e., propagation likely). Therefore there is a 50% chance that an incident edge will be removed when a node is removed. (While it is not strictly correct, reversal of flow is considered relatively unlikely in this propagation model because of the very local nature of reactive power, which supports bus voltage. This non-reversal simplification is not expected to change results greatly in the context of this study.)

The use of a directed failure propagation model for edges should not be viewed as inconsistent with the undirected model for nodes. Electric networks are only directed graphs for *a particular configuration*, and the removal of a component usually causes changes in flow, but not always reversals. However, over the entire system, a change in flow direction on one edge does not change the aggregate probabilities for adjacent nodes. For example, if two nodes are connected by an edge, then one receives power and the other supplies it. A flow reversal simply switches which is which, with no overall effect on the system aggregate probabilities.

Therefore, when we combine the two probabilities, we obtain the overall probability of a node removal propagating via a single edge to another node:

$$P\{\to 1\} = \frac{\rho_1}{2\gamma}. \qquad (6)$$

The probability of a node removal propagating to a node two edges away is simply $P\{\to 2\} = P\{\to 1\}^2$ and the probability of a node removal propagating to a node $i$ edges away is

$$P\{\to i\} = P\{\to 1\}^i = \left(\frac{\rho_1}{2\gamma}\right)^i. \qquad (7)$$

Integrating the product of this probability with the connectivity probabilities for all *N* nodes, we obtain the aggregate probability of all system failure states, which is equivalent to the system LOLP. Thus, we have

$$LOLP = \int_1^N \rho(x) P\{\to x\} dx = \int_1^N \rho_1 x^{-\lambda} \left(\frac{\rho_1}{2\gamma}\right)^x dx. \qquad (8)$$

Comparisons of this result for both major North American systems with estimates of other systems made by Melo *et al.* [8] are shown in Table 3.

---

[a] For a system of *N* nodes, the probability of failure propagation from link to node is slightly lower: $P\{\to node\} = \int_1^N \rho_1 x^{-\gamma-1} dx = \frac{1}{\gamma} \rho_1 (1 - N^{-\gamma})$

Table 3: Comparison of analytic results to estimates by Melo *et al.* [8]

| System | LOLP | $\sigma_{LOLP}$ |
|---|---|---|
| Western System | 0.026 | +0.0056<br>-0.0046 |
| Eastern Interconnect | 0.026 | +0.0039<br>-0.0034 |
| Bonneville Power Administration (Melo) | 0.027 | - |
| Modified IEEE Reliability Test System[a] (Melo) | 0.043 | - |
| Brazilian Southern/South-Eastern (Melo) | 0.047 | - |

[a] In 1986, a number of modifications to the 1979 Reliability Test System (RTS-79) were introduced by the Institute of Electrical and Electronics Engineers, Inc. (IEEE) to expand the data related generation systems [see Pinhero *et al*. "Probing the new IEEE Reliability Test System (*RTS-96*): HL-II Assessment", *IEEE Transactions on Power Systems* **13** 1 (1993)].

## 5. Discussion

The closest correlation is with the estimate by Melo *et al.* [8] for the Bonneville Power Administration (BPA) system, which is in fact a part of the Western System considered above. We can only speculate regarding the differences between the results for the Modified IEEE Reliability Test System (RTS-79) and the Brazilian Southern/South-Eastern system (SSE). One possibility is that the RTS-79 and SSE system are intrinsically less reliable because of the different nodal contribution of equipment reliability, an aspect of overall system reliability that is treated more explicitly by Melo *et al*. than by the proposed model. Another intriguing possibility that should be considered is that these two systems are less scale-free because they are not as large and/or did not evolve over sufficient time so that self-organizing phenomena do not occur to the same degree as in the North American grids. Ultimately we recommend examining the RTS-79 and SSE system using the proposed model to uncover the degree to which this is the case.

Comparing these results to the previous studies we believe the increase in the $\gamma$ value is attributable to two differences in the models used. First, the way in which leaf nodes are counted may increase the slope somewhat. Second, and perhaps more significantly, we believe that inclusion of all network nodes in model regardless of voltage greatly increases the range of scales over which the statistics are collected. The total number of nodes we considered is 314,123, compared to the 14,099 in the Albert model, and the 4,941 in the Watts and Strogatz. It is reasonable to expect that lower voltage nodes on the system reach capacity constraints for lower values of degree *k* than do higher voltage nodes. This would tend to increase the value of $\gamma$ when the $\kappa$ cut-off term is ignored, in spite of any attempt to ignore larger degree nodes, as in this study.

It is useful to reflect on the relationship of generating functions discussed in Newman *et al*. [12] to the proposed failure propagation model. As we have seen, the aggregate electric network in an operational configuration is essentially an aggregate directed graph like the "bow-tie" configuration proposed by Broder *et al*. [13], such that generators are the links in, loads are the links out, and the transmission system is the strongly connected component. Generating functions such as those discussed by Newman *et al.* are very useful in estimating many properties of networks, such as the *n*th moment. For example, mean propagation path length can be evaluated using such functions if the network structure has been properly characterized. We urge caution before generalizing such an approach to all electric networks. As the results above suggest, less mature networks appear to have greater multi-node outage probabilities than expected for scale-free networks. So while the temptation to use such methods may be great, the potential for less mature power networks to have non-scale-free structure can lead to erroneous conclusions.

---
[a]

Studies of North American electric system blackout events over the last three decades by Carreras *et al.* [14] and Chen *et al.* [15] reveal a "fat-tail" probability distribution for both the size and frequency of system outages. However, these phenomenological power laws are not *per se* evidence of a scale-free network, nor is it obvious that a scale-free network structure will necessarily lead us to predict a power law outage probability. The results of our analysis only clarify the relationship between overall outage probability as measured by the LOLP and the most general scale-free aspects of the network structure. Using this, one can only relate the quantitative system failure data gathered by electric grid operators to general topological characteristics of the electric network itself. But we have yet to uncover whether this result does indeed lead to outage size and frequency probabilities consistent with previous observations.

## 6. Conclusion

We have confirmed the accuracy of the Barabási-Albert scale-free network model for two North American electric transmission systems using only the most general topological data about bulk electric power transmission networks. We proposed a simple model of failure propagation in the electric power grid based on the known physical and control laws in the systems, which we combined with the scale-free network connectivity probability distribution to estimate the loss-of-load probability (LOLP) reliability index for two major North American electric grids. The LOLP estimates based on the Barabási-Albert network model were compared to other LOLP estimates for the Bonneville Power Administration's region of the western grid and other similar electric systems. We quantitatively confirm that the measured reliability of this part of the North American electric grids closely matches the reliability predicted using the Barabási-Albert model.

We recommend further examination of grid reliability models using Barabási-Albert models be conducted with physically accurate models of failure propagation to refine these results and resolve the remaining inconsistencies with previous analyses, particularly with respect to the $\gamma$ power-law coefficient. In addition, this study did not attempt to confirm previously reported values of the exponential cut-off coefficient $\kappa$, which arises from cost and capacity constraints. Finally, the present analysis was made using only two of many available reliability models for the eastern and western grids and should be repeated for a wider variety of operating conditions and times of year.

**Acknowledgements**

This work was supported by the Energy Systems Transformation Initiative at the U.S. Department of Energy's Pacific Northwest National Laboratory. Pacific Northwest National Laboratory is operated by Battelle Memorial Institute for the U.S. Department of Energy under Contract DE-AC05-76RL01830.